\documentclass[prd,twocolumn,nofootinbib,preprintnumbers,showpacs]{revtex4}

\newcommand\lsim{\mathrel{\rlap{\lower4pt\hbox{\hskip1pt$\sim$}}
   \raise0.2pt\hbox{$<$}}}
\newcommand\gsim{\mathrel{\rlap{\lower4pt\hbox{\hskip1pt$\sim$}}
   \raise0.2pt\hbox{$>$}}}
\usepackage{epsfig,graphicx}

\def\cosb{\cos\beta}
\def\sinb{\sin\beta}

\def\mw{{M_W}}

\def\sqrtwo{\sqrt{2}}

\def\chsnsn{C_{\higgsi\snr\snr}}

\def\chjsnsn{C_{\higgsj\snr\snr}}

\def\l{\lambda}
\def\k{\kappa}
\def\vevs{v_s}

\def\ln{{\lambda_N}}
\def\aln{{A_{\lambda_N}}}
\def\an{{A_{N}}}

\def\neutmassone{{m_{{\tilde\chi}_{1}}}}

\def\snr{{\tilde N_1}}

\def\snmass{{m_{\tilde N}}}

\def\snmassr{{m_{\tilde N_1}}}

\def\rhnmass{{m_{N}}}

\def\higgsi{{H_i^0}}
\def\higgsj{{H_j^0}}

\def\higgsone{{H_1^0}}

\def\hmassone{m_{H_1^0}}
\def\hmasstwo{m_{H_2^0}}
\def\hmassthree{m_{H_3^0}}

\def\hmassi{m_{H_i^0}}

\def\hcompid{S_{H_i^0}^1}
\def\hcompiu{S_{H_i^0}^2}
\def\hcompis{S_{H_i^0}^3}
\def\hcompij{S_{H_i^0}^j}

\def\phmassone{{m_{A_1^0}}}
\def\phmasstwo{{m_{A_2^0}}}

\def\chmass{{m_{H^+}}}

\newcommand{\crosssec}{\sigma_{\snr-p}^{\rm SI}}
\newcommand{\nmh}{{\tt NMHDECAY}}
\newcommand{\snrelic}{{\Omega_\snr h^2}}

\begin{document}

\title{Right-handed sneutrino as thermal dark matter 
}

\author{ David G.~Cerde\~no ${}^{1,2}$, Carlos Mu\~noz ${}^{1,2}$, 
 and Osamu Seto ${}^2$ }

\affiliation{
${}^1$ Departamento de F\'{\i}sica Te\'{o}rica C-XI, and 
 ${}^2$ Instituto de F\'{\i}sica Te\'{o}rica UAM-CSIC, 
 Universidad Aut\'{o}noma de Madrid, Cantoblanco, E-28049, Madrid, Spain}


%
\begin{abstract}
  We study an extension of the MSSM with a singlet supermultiplet 
  $S$ with coupling 
  $SH_1H_2$ in order to solve the
  $\mu$ problem as in the NMSSM, and right-handed neutrino
  supermultiplets 
  $N$ with couplings
  $SNN$ in order to generate dynamically electroweak-scale Majorana masses. 
  We show how in this model a purely right-handed sneutrino can be a
  viable candidate for cold dark matter in the Universe.
  Through the direct coupling to the singlet, the
  sneutrino can not only be thermal relic dark matter but
  also have a large enough scattering cross section with nuclei to
  detect it directly in near future, in contrast with most 
  other right-handed sneutrino dark matter models. 
\end{abstract}
\pacs{PACS: 95.35.+d, 12.60.Jv, 98.80.Cq}
\preprint{FTUAM-08/12, IFT-UAM/CSIC-08-47} 
\date{\today}
\maketitle


\section{Introduction.}

Weakly interacting massive particles (WIMPs) are among the 
best motivated candidates for explaining the cold dark matter in 
the Universe. WIMPs appear in many interesting 
extensions of the standard model 
providing new physics at the TeV scale. 
Such is the case of supersymmetric models, in which 
imposing a discrete symmetry (R-parity) to avoid
rapid proton decay renders the lightest supersymmetric particle (LSP)
stable and 
thus a good dark matter (DM) candidate.

The minimal supersymmetric extension of the 
standard model
(MSSM) provides two natural candidates for WIMPs, the 
neutralino and
the (left-handed) sneutrino. 
The neutralino is a popular and extensively studied
possibility.  
On the contrary, the left-handed sneutrino in the
MSSM~\cite{Ibanez} is not a viable dark
matter candidate. Given its sizable 
coupling to the $Z$ boson, they either annihilate too rapidly,
resulting in a very small relic abundance, or give rise to a large
scattering cross section off nucleons and are excluded by direct DM
searches~\cite{Falk:1994es} 
(notice however that 
the inclusion of a lepton number violating operator can reduce the
detection cross
section~\cite{Hall:1997ah}).

However, there is a strong motivation to consider an extension of the
MSSM, the fact that neutrino oscillations imply tiny but non-vanishing
neutrino masses.  
These can be obtained introducing right-handed neutrino superfields.
Several models have been proposed to revive sneutrino DM 
by reducing its coupling with Z-boson. This can be achieved by
introducing 
a mixture of left- and right-handed
sneutrino~\cite{ArkaniHamed:2000bq,Arina:2007tm,valle}, 
or by considering a purely right-handed
sneutrino~\cite{Asaka:2005cn,Gopalakrishna:2006kr,McDonald:2006if,Lee:2007mt}.
In the former, 
a significant left-right mixture is realized  
by adopting some particular supersymmetry
breaking with a large trilinear term~\cite{ArkaniHamed:2000bq}. 
Such a mechanism is not available in the standard
supergravity mediated supersymmetry breaking, where trilinear terms are 
proportional to the small neutrino Yukawa couplings.
Recently, another realization of large mixing was pointed out~\cite{valle} 
by abandoning the canonical see-saw formula for neutrino
masses. 
On the other hand, 
pure right-handed sneutrinos cannot be thermal relics, since their
coupling to ordinary matter is extremely reduced by the neutrino Yukawa 
coupling~\cite{Asaka:2005cn,Gopalakrishna:2006kr,McDonald:2006if}. 
Furthermore, these
would be unobservable in direct detection experiments. 
Another possibility to obtain the correct thermal relic density would
consist of coupling the right-handed sneutrino to the observable
sector, e.g., via an extension of the gauge
\cite{Lee:2007mt} or Higgs \cite{pilaftsis,dp} sectors

There is one more motivation to consider another
extension of the MSSM, 
the so-called ``$\mu$ problem''~\cite{Kim:1983dt}.
The superpotential in the MSSM contains a bilinear term, 
$\mu H_1
H_2$.
Successful radiative electroweak symmetry breaking (REWSB) 
requires $\mu$ of the order of the electroweak scale. 
The next-to-minimal supersymmetric standard model (NMSSM) 
offers a simple solution 
by 
introducing a singlet superfield $S$
and
promoting the bilinear term to a trilinear coupling $\l S H_1 H_2$. 
After REWSB, 
$S$ develops a vacuum expectation value (VEV)
of order of the electroweak scale
thereby providing an effective $\mu$ term, $\mu=\lambda \langle S\rangle$.
Furthermore, the NMSSM also alleviates the 
``little hierarchy problem'' of the Higgs sector in 
the MSSM~\cite{BasteroGil:2000bw} and has an attractive
phenomenology, featuring light Higgses and interesting consequences
for neutralino DM \cite{Cerdeno:2004xw}. 
Although the $Z_3$ symmetry
of the NMSSM may give rise to a cosmological domain wall problem, 
this can be avoided with the inclusion of
non-renormalisable operators.

Motivated by the above two issues, we study an extension of the MSSM
where singlet scalar superfields are included 
\cite{ko99,pilaftsis}. 
A singlet $S$  in order 
to solve the $\mu$ problem as in the NMSSM (and which accounts for extra
Higgs and neutralino states)
and right-handed
neutrinos $N$ to obtain non-vanishing neutrino Majorana masses with
the canonical, but low scale, see-saw mechanism. 
Terms of the type $SNN$ in the superpotential can generate dynamically
Majorana masses 
through the VEV of the singlet $S$. 
In addition, 
the presence of right-handed sneutrinos, $\tilde N$, 
with a weak scale mass
provides a new possible DM candidate within the WIMP category. 

In this letter we analyse the properties of right-handed sneutrinos,
showing that not only 
they can be thermally produced in sufficient amount to
account for 
the DM in the Universe because of the direct coupling between
$S$ and $N$, but also that their elastic scattering cross
section off nuclei is large enough to allow their detection in future
experiments.

\section{The Model.}

The superpotential in our construction is an extension of that of the
NMSSM, including new trilinear coupling among the singlets $S$ and $N$
and Yukawa terms to provide neutrino masses. It reads
\begin{eqnarray}
  W &=& W_{\rm NMSSM} + \lambda_N S N N + y_N H_2 \cdot L N,  \\
  W_{\rm NMSSM} &=& Y_u H_2 \cdot Q u + Y_d H_1 \cdot Q d + Y_e H_1 
  \cdot L e \nonumber \\ 
  &&  -\lambda S H_1 \cdot H_2 + \frac{1}{3}\kappa S^3 ,
\end{eqnarray}
where flavour indices are omitted and the dot denotes the $SU(2)_L$
antisymmetric product. As in the NMSSM, a global $Z_3$ symmetry is
imposed, 
so that there are no supersymmetric mass
terms in the superpotential.  Note that the term $NNN$ and $SSN$ are gauge
invariant but not consistent with R-parity and thus are not included. 
We also assume no CP violation in the Higgs sector.

Once REWSB takes place and the Higgs fields take non-vanishing 
VEVs, 
$(v_{1, 2}, v_s) = (\langle H_{1, 2}\rangle, \langle S \rangle)$, an
effective Majorana mass term in the neutrino sector 
is generated, $M_N = 2 \ln v_s$.
Light masses for left-handed neutrinos are then obtained via a see-saw
$ m_{\nu_L} = {y_N^2 v_2^2}/{M_N}$, 
which implies Yukawa couplings $y_N\lesssim{\cal O} (10^{-6})$ of the
same order of the electron Yukawa.

The sneutrino mass matrix can be read from the quadratic terms in the
scalar potential as 
\begin{eqnarray}
  && 
  \frac{1}{2}(\tilde\nu{_L}_1, \tilde{N}_1)
  \left(
  \begin{array}{cc}
    m_{L\bar{L}}^2           &  m_{L\bar{R}}^2+m_{LR}^2 \\
    m_{L\bar{R}}^2+m_{LR}^2  &  m_{R\bar{R}}^2 + 2m_{RR}^2 \\
  \end{array}
  \right)
  \left(
  \begin{array}{c}
    \tilde\nu{_L}_1  \\
    \tilde{N}_1 \\
  \end{array}
  \right)+ \nonumber \\
  &&  
  \frac{1}{2} (\tilde\nu{_L}_2, \tilde{N}_2)
  \left(
  \begin{array}{cc}
    m_{L\bar{L}}^2           &  m_{L\bar{R}}^2-m_{LR}^2 \\
    m_{L\bar{R}}^2-m_{LR}^2 &  m_{R\bar{R}}^2 - 2m_{RR}^2 \\
  \end{array}
  \right)
  \left(
  \begin{array}{c}
    \tilde\nu{_L}_2  \\
    \tilde{N}_2 \\
  \end{array}
  \right) .
  \label{SneutrinoMass}
\end{eqnarray}
Here, sneutrinos are decomposed in real and imaginary components as
${\tilde\nu_L} \equiv (\tilde\nu{_L}_1 + i
\tilde\nu{_L}_2)/\sqrt{2}$  and
$\tilde{N} \equiv (\tilde{N}_1 + i \tilde{N}_2)/\sqrt{2}$, 
and all parameters are defined by
\begin{eqnarray}
  m_{L\bar{L}}^2
  &\equiv& m_{\tilde{L}}^2 + |y_N v_2|^2 + {\rm D-term} , \nonumber \\
  m_{LR}^2
 &\equiv& y_N\left(-\lambda_N \vevs v_1 \right)^{\dagger}
  + y_N A_N v_2 , \nonumber\\
  m_{L\bar{R}}^2
  &\equiv& y_N v_2 \left(-\lambda_N \vevs \right)^{\dagger} , \nonumber\\
  m_{R\bar{R}}^2 
  &\equiv& m_{\tilde{N}}^2 +|2\lambda_N \vevs|^2 
  + |y_N v_2|^2 , \nonumber\\
  m_{RR}^2
  &\equiv& \lambda_N \left( A_{\lambda_N} \vevs
  + (\kappa \vevs^2-\lambda v_1 v_2)^{\dagger}  \right) ,
  \label{sn:mrr}
\end{eqnarray}
where $m_{\tilde{L}}^2$, $m_{\tilde{N}}^2$, $\aln$, and $\an$, 
are the new soft parameters. 
These are assumed to be real for simplicity, 
so that the real and imaginary parts of sneutrinos do not mix.
The mixing between left- and right-handed sneutrinos, 
induced by $m_{LR}^2$ and $m_{L\bar{R}}^2$, is 
proportional
to the small neutrino Yukawa coupling $y_N$,
and therefore negligible. 
Note that $m_{RR}^2$ splits the masses of $\tilde{N}_1$ and $\tilde{N}_2$. 
$\tilde{N}_2$ is lighter than $\tilde{N}_1$ for $m_{RR}^2 > 0$ and vice versa.

Although the right-handed sneutrino may have a non vanishing
VEV breaking R-parity spontaneously~\cite{ko99}, 
by solving the stationary condition 
we find that the origin $\tilde{N} =0$ 
is the true minimum if $m_{R\bar{R}}^2- 2|m_{RR}^2|  > 0$, 
which is precisely the condition for the
lightest right-handed sneutrino mass squared~(\ref{SneutrinoMass}) 
to be positive.
Hereafter we only consider cases where this condition is satisfied.
In such a case, 
the Higgs potential coincides with 
that in the NMSSM.

The coupling between a Higgs boson, $\higgsi$, 
and two right-handed sneutrinos determines most of the sneutrino
phenomenological properties. 
It can be calculated
from the superpotential and Lagrangian and reads
\begin{eqnarray}
  \chsnsn&=&
  \frac{2\l\ln\mw}{\sqrtwo g}
  \,\left(\sinb\hcompid+\cosb\hcompiu\right)\, + \nonumber\\
  &&\left[
    \left(4\ln^2+2\k\ln\right)\vevs+\frac{\ln\aln}{\sqrtwo}
    \right]\hcompis \,,
  \label{csnsn} 
\end{eqnarray}
where $\hcompij$ ($j=1,2,3$) are the elements of the Higgs
diagonalisation matrix.

\section{Thermal relic density.}
The right-handed sneutrino, having a mass of order the EW scale, can
be the LSP in our construction for adequate choices of the input
parameters (in particular, for small $\snmass$). In such a case, it
constitutes a good candidate for DM. 
In order to determine its viability, its
thermal relic abundance, $\snrelic$, 
needs to be calculated and compared to the
WMAP result,  $0.1037\le\Omega h^2\le 0.1161$~\cite{wmap5yr}.
The possible products for
$\snr\snr$ annihilation include
\begin{itemize}\itemsep -0.5ex
\item 
  $W^+\, W^-$, $Z\, Z$, and $f \bar{f}$
  via $s$-channel Higgs exchange;
\item
  $H_i^0\, H_j^0$, 
  via $s$-channel Higgs exchange, $t$-
  and $u$-channel sneutrino exchange, and a scalar quartic coupling;
\item
  $A_a^0\, A_b^0$, and $H_i^+\, H_j^-$,  
  via  $s$-channel Higgs exchange, and a scalar quartic coupling;
\item
  $Z\,A_a^0$ and $W^\pm\,H^\mp$
  via $s$-channel Higgs exchange;
\item
  $NN$, 
  via $s$-channel Higgs exchange and via $t$- and $u$-channel
  neutralinos exchange.
\end{itemize}
The processes suppressed by the neutrino Yukawa $y_N$, e.g., those
involving $Z$ exchange along an $s$-channel, are negligible
and have not been taken into account. 
Notice that similar models where the Higgs sector is extended share
some of these channels \cite{dp}. 
It is obvious that the annihilation cross section is very dependent 
on the structure of the Higgs sector. In particular, all the processes
involve 
$s$-channel Higgs exchange, which implies the presence of rapid
annihilation in the resonances, when $2\snmassr\approx \hmassi$. 
In addition, annihilations into a neutral Higgs pair turn
out to be one of the dominant channels, implying a significant
decrease in $\snrelic$ when $\snmassr>\hmassi$.
This is interesting, since very light
Higgses are possible (as long as they have a significant singlet
component) in the NMSSM.
Another important contribution comes from the annihilation into a pair of
right-handed neutrinos when $\snmassr>\rhnmass$.

In our calculation we do not include 
coannihilation effects. 
For our choice of parameters and in wide regions of the NMSSM 
these are only important in the regions 
in which the LSP changes from sneutrino to neutralino and do not
affect our conclusions.

Our input parameters are, on the one hand, 
the usual NMSSM degrees of freedom, 
$\lambda, \, \kappa,\, \tan \beta,\, \mu,\, A_\lambda, \, A_\kappa$, 
which we define at low-energy. 
Regarding the soft parameters, 
we assume that gaugino masses mimic, at
low-energy,
the values obtained from a hypothetical GUT unification. 
Low-energy observables, such as the muon anomalous magnetic moment and 
BR($b\to s\gamma$), pose stringent constraints on the NMSSM parameter
space. 
In order to avoid these, we consider an example with 
$m_{L,E}= 150$~GeV, $m_{Q,U,D}= 1000$~GeV,
$M_1=160$~GeV, $A_E=-2500$~GeV $A_{U,D}=2500$~GeV,
$A_\lambda=400$~GeV, $A_\kappa=-200$, $\mu=130$~GeV and 
$\tan\beta=5$, that was studied in~\cite{Cerdeno:2004xw} (see Fig.7
there). 
The choice $\lambda=0.2$ and $\kappa=0.1$ corresponds to a very
characteristic point of the NMSSM, 
featuring a very light Higgs
mass, $\hmassone=59.7$~GeV (compatible with LEP bounds due to its
large singlet component),
and a lightest neutralino with $\neutmassone=88.5$~GeV (which sets
the upper limit for $\snr$ as the LSP). The masses of the remaining
two scalar Higgses are
$\hmasstwo=125.6$~GeV and $\hmassthree=550.8$~GeV.
The pseudoscalar masses in this example are $\phmassone=199.6$~GeV and 
$\phmasstwo=548.9$~GeV, where the lightest pseudoscalar also has a
significant singlet composition. 
Finally, the resulting mass of the charged Higgs is
$\chmass=553.7$~GeV. 
The viability of this set of NMSSM parameters is checked with
the \nmh~2.0 code~\cite{Ellwanger:2005dv}, based on which
we have built a package which
calculates the sneutrino relic density using the numerical procedure
described in \cite{relicd}.

\begin{figure}
  \epsfig{file=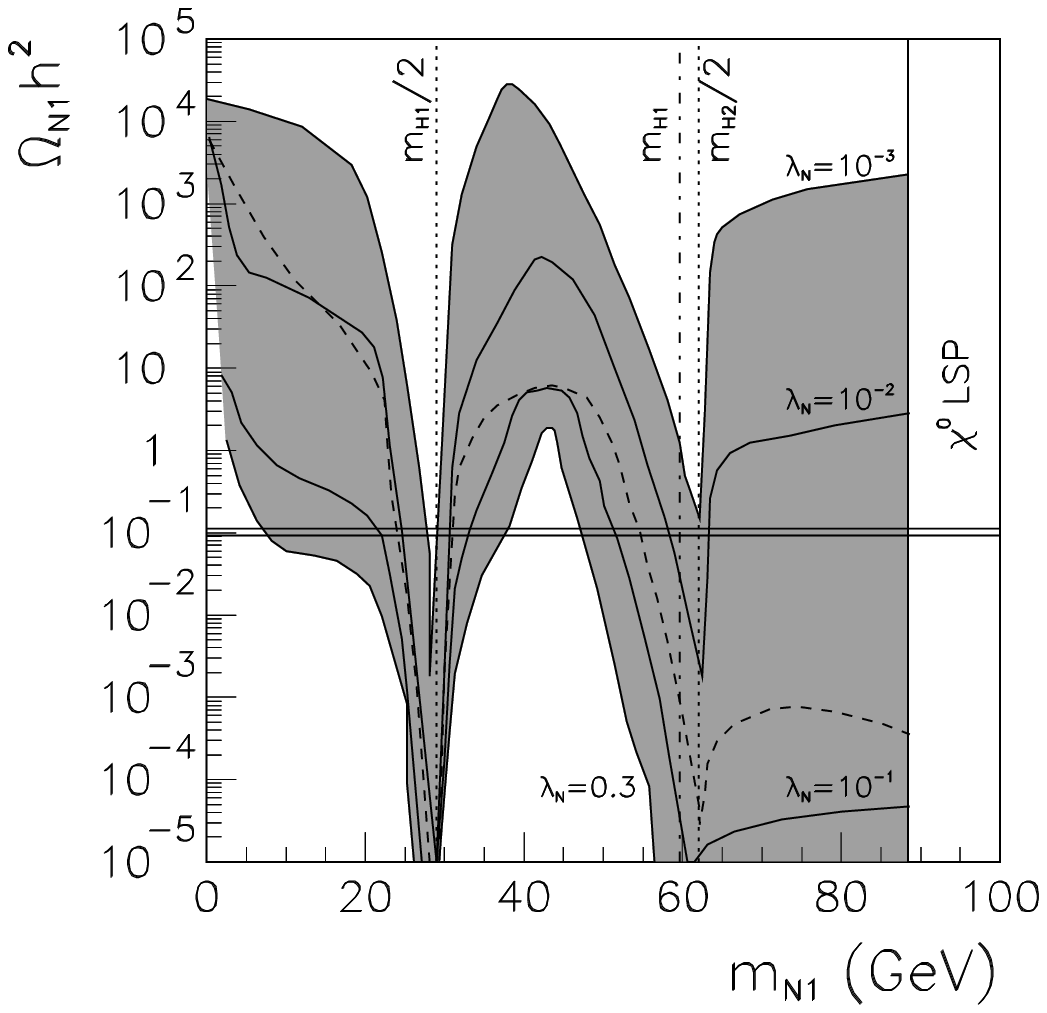,width=9.cm}
  \vspace*{-1cm}
  \caption{$\snrelic$ as a function of $\snmassr$ for
    $\ln\subset[10^{-3},0.3]$  (grey
    area).  
    The vertical dotted lines indicate the location of the 
    Higgs resonances for $2\snmassr\approx m_{H_{1,2}^0}$, and the
    dot-dashed line indicates the opening of the annihilation channel
    into $\higgsone\higgsone$. 
    Points below the dashed line have $\snmassr<\rhnmass$.
    The
    vertical solid line represents the value of the lightest
    neutralino mass. 
  }
  \label{fig:relic}
\end{figure}

Our model contains
three new parameters to be fixed, 
$\lambda_N,\, \snmass,\, \aln$.
In order to illustrate the theoretical predictions for 
$\snrelic$ we set  $\aln=250$~GeV and vary $\ln$ and
$\snmass$ in the ranges
$[10^{-3},0.3]$ and $[0,200]$~GeV, respectively, 
excluding those points in which $\snr$ is not the LSP or is
tachyonic. 
The resulting $\snrelic$ is shown in
Fig.\,\ref{fig:relic}, where
the large suppression on the Higgs resonances for the two lightest
Higgses 
is clearly evidenced (the resonance corresponding to 
the heaviest scalar Higgs would be
located at $\hmassthree/2\approx275$~GeV and is therefore not
accessible in the region where the sneutrino is the LSP in the present
example).
The relic abundance increases as $\ln$ decreases due to the reduction
in $\chsnsn$. 
Remarkably, the correct relic density can be obtained with natural
values of $\ln$. In particular, when annihilation into Higgses is
possible ($\snmassr>\hmassone$), one needs 
$10^{-2}\lsim\ln\lsim 10^{-1}$. 
Notice also that very light
$\snr$ are viable with $\ln\gsim10^{-1}$ through the annihilation into 
$b\bar b$. 
For our choice of parameters a lower bound
$\snmassr\approx 10$~GeV is obtained.

\section{Direct detection.}

The direct detection of sneutrinos would take place through their
elastic scattering with nuclei inside a DM detector. At the
microscopic level, the
low-energy interaction of sneutrinos and quarks 
can be described 
by an effective
Lagrangian. In our case, there is only one contribution (at
tree level) to this process, the $t$-channel exchange of
neutral Higgses. 
In
terms of the Higgs-sneutrino-sneutrino coupling, one can write
\begin{equation} 
  {\cal L}_{eff}\supset \sum_{j=1}^3\frac{\chjsnsn
    Y_{q_i}}{m_{H_j^o}^2}
  \tilde N\tilde N
  \bar q_i q_i
  \equiv
  \alpha_{q_i} 
  \tilde N\tilde N
  \bar q_i q_i\,,
\end{equation}
where 
$Y_{q_i}$ is the
corresponding quark Yukawa coupling and $i$ labels up-type quarks
($i=1$) and down-type quarks ($i=2$). The effective
Lagrangian contains no axial-vector coupling since the sneutrino is a
scalar field, thus implying a vanishing 
spin-dependent cross section.

\begin{figure}
  \epsfig{file=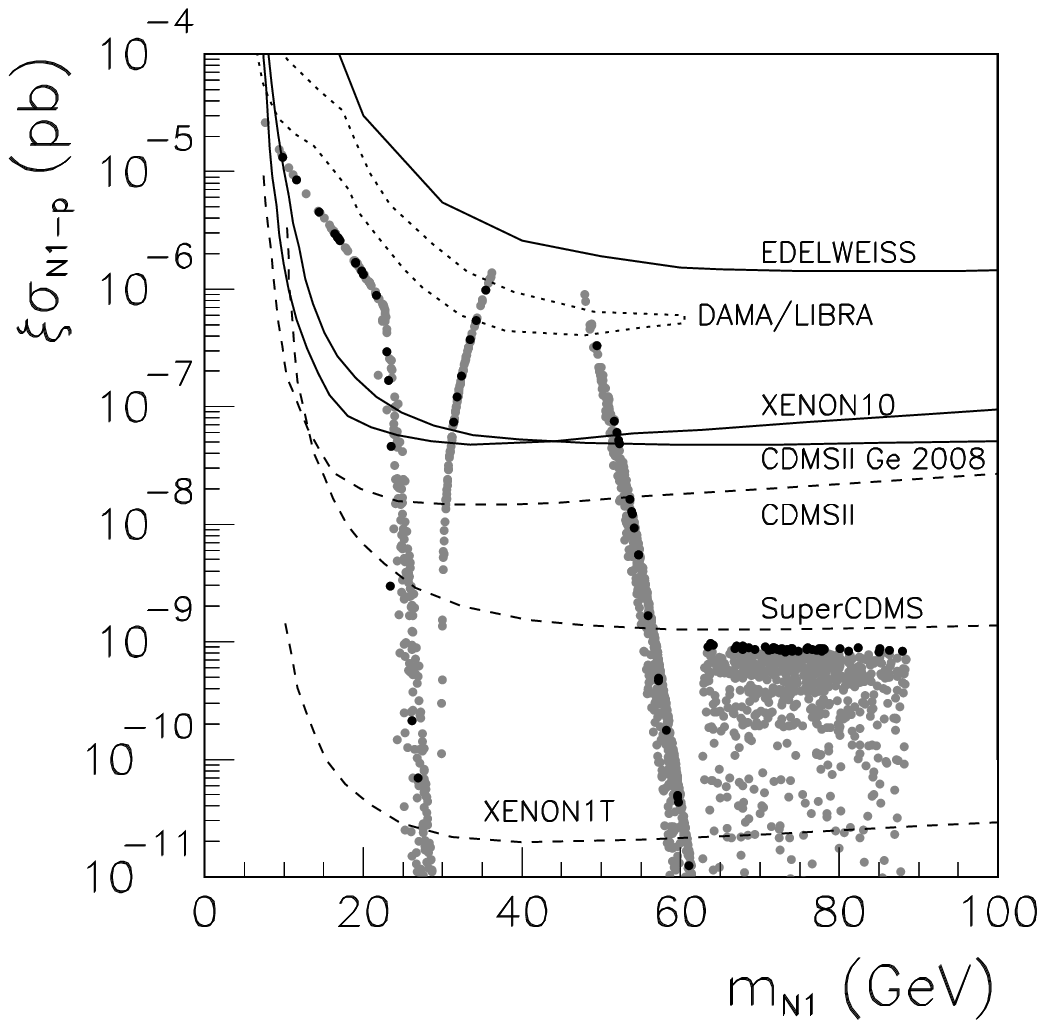,width=9.cm}
  \vspace*{-1cm}
  \caption{Theoretical predictions for
    $\xi\crosssec$, as a function of $\snmassr$. 
    The
    sensitivities of present and projected experiments are represented
    by means of solid and dashed lines, respectively, in the case of an
    isothermal spherical halo. The area bounded by
    dotted lines is consistent with the interpretation of the
    DAMA/LIBRA \cite{libra}
    experiment in terms of a WIMP.}
  \label{fig:sncross}
\end{figure}

The total spin-independent sneutrino-proton 
scattering cross section yields
\begin{equation}
  \crosssec = \frac1\pi\frac{m_p^4}{(m_p+\snmassr)^2}\,f_p^2\,,
\end{equation}
where $m_p$ is the proton mass and the expression for $f_p$ in terms
of the hadronic matrix elements parametrizing the quark content
of the proton
can be found, e.g., in
\cite{eos08}.

The scattering
cross section is also very dependent on the features of the Higgs
sector. In particular, $\crosssec$ becomes larger when
$\chsnsn$ (\ref{csnsn}) increases (e.g.,
when $\l$, $\ln$ or $\aln$ are enhanced) 
and/or the mass of the lightest CP-even Higgs decreases.

The theoretical predictions for
$\xi\crosssec$  are represented as a function of the sneutrino mass in
Fig.\,\ref{fig:sncross}.
The sneutrino fractional density $\xi$, 
is defined to be
$\xi={\rm min}[1,\snrelic/0.1037]$ in order to have a
rescaling of the signal for 
subdominant DM in the halo~\cite{rescaling}.
Black dots correspond to points with a relic density consistent with
the WMAP results, whereas grey dots stand for those with
$\snrelic\le0.1$ in which $\snr$ is subdominant.

The right-handed sneutrino in our model
is not yet excluded by direct searches for DM.
Interestingly, the predicted $\crosssec$ lies within the reach of
projected detectors, such as SuperCDMS and XENON1T
(unlike a pure right-handed sneutrino with only Yukawa interactions).  

Regarding the possible indirect detection of this WIMP candidate,
notice that not being a Majorana fermion, 
contrary to the neutralino case, sneutrino annihilation
into a $f\bar f$ pair would not be helicity suppressed. 
Thus it can potentially lead to larger signals 
(as in the case of Kaluza-Klein dark matter).
A detailed analysis of the detectability of these signals will be
presented elsewhere.

\vspace*{0.5cm}
\section{Conclusions.}

We propose the right-handed sneutrino as a viable thermal DM candidate in an
extension of the MSSM where the singlet superfields,
$S$ and $N$, are included to solve the $\mu$ problem and account for
neutrino masses. A direct coupling between $S$ and $N$ provides a 
sufficiently large annihilation cross section for the right-handed
sneutrino, as well as a 
detection cross section in the range of future direct DM
searches.

DGC was supported by the program 
``Juan de la Cierva''. 
CM was supported by the MEC project
FPA2006-05423 and 
the EU  program
MRTN-CT-2004-503369.
DGC  and
CM were also supported by the 
MEC project FPA2006-01105, 
and  the EU 
network
MRTN-CT-2006-035863. 
OS was supported by the MEC project FPA 2004-02015.
We thank the ENTApP Network of the
ILIAS project RII3-CT-2004-506222 and 
the project HEPHACOS P-ESP-00346 
of the Comunidad de Madrid.


\end{document}